\documentclass[submission,copyright,creativecommons]{eptcs}
\providecommand{\event}{TARK 2017} 
\usepackage{breakurl}             
\usepackage{underscore}           

\usepackage{graphicx}
\usepackage{caption}
\usepackage{subcaption}
\usepackage{amsthm}
\usepackage{amsfonts}
\usepackage{amsmath}

\newtheorem{theorem}{Theorem}
\newtheorem{definition}{Definition}
\newtheorem{axiom}{Axiom}
\newtheorem{proposition}{Proposition}
\newtheorem{example}{Example}

\newcommand{\commentout}[1]{}

\makeatletter
\providecommand*{\cupdot}{%
 \mathbin{%
 \mathpalette\@cupdot{}%
 }%
}
\newcommand*{\@cupdot}[2]{%
 \ooalign{%
 $\m@th#1\cup$\cr
 \hidewidth$\m@th#1\cdot$\hidewidth
 }%
}
\makeatother

\title{Group Recommendations:\\Axioms, Impossibilities, and Random Walks}
\author{Omer Lev
\institute{University of Toronto\\ Canada}
\email{omerl@cs.toronto.edu}
\and
Moshe Tennenholtz
\institute{Technion\\
Israel}
\email{moshet@ie.technion.ac.il}
}
\def\titlerunning{Group Recommendations: Axioms, Impossibilities, and Random Walks}
\def\authorrunning{Lev \& Tennenholtz}

\begin{document}
\maketitle

\begin{abstract}
We introduce an axiomatic approach to group recommendations, in line of previous work on the axiomatic treatment of trust-based recommendation systems, ranking systems, and other foundational work on the axiomatic approach to internet mechanisms in social choice settings. In group recommendations we wish to recommend to a group of agents, consisting of both opinionated and undecided members, a joint choice that would be acceptable to them. Such a system has many applications, such as choosing a movie or a restaurant to go to with a group of friends, recommending games for online game players, \& other communal activities.

Our method utilizes a given social graph to extract information on the undecided, relying on the agents influencing them. We first show that a set of fairly natural desired requirements (a.k.a axioms) leads to an impossibility, rendering mutual satisfaction of them unreachable. However, we also show a modified set of axioms that fully axiomatize a group variant of the random-walk recommendation system, expanding a previous result from the individual recommendation case.
\end{abstract}
\maketitle
\section{Introduction}

Since the rise of the internet, we have seen reputation systems, ranking systems, trust systems, recommendation systems, affiliate marketing in social networks, and more, flowering in its midst. This recent wave of online social systems is typically associated with a large amount of data that is collected online which leads to the ``big data'' approach to the utilization of such information. Quite surprisingly, however, the abundance of available data does not help system designers come up with the right design for online systems in the first place. Indeed, available data is typically generated by the use of a particular system, and mining the data generated by users while interacting with one system does not provide a tool for exploring the overwhelmingly large design space. Interestingly, the main practical approach of software and hardware design, the formal specification of clear system requirements and the implementation of a system satisfying these exact requirements, has not been used often.
This classical approach, when adapted to the context of multi-agent systems, coincides with extensions of a famous tool of social choice theory and cooperative game theory, namely {\em the axiomatic approach}.

Perhaps the best known axiomatic theory in the social sciences is the theory of social choice~\cite{Plo76}. In that setting we have a set of voters and a set of alternatives, where each voter has a ranking over the set of alternatives, and our aim is to find a good aggregation of the individual rankings into a global ranking.
Various properties of such aggregation functions have been considered and have led to various characterizations of particular systems as well as impossibility results showing no system can satisfy certain sets of properties all at once~\cite{Arr51}. In the internet setting, the study of ranking and reputation systems, such as page-ranking systems~\cite{AT05b}, defines a natural extension of classical social choice to the case where the set of voters and the set of alternatives coincide. In such a setting the classical axioms of social choice become less relevant, and are replaced with new axioms which lead to completely new theory. Further removed from direct extensions of social choice, one can find systems originating from personalized versions of ranking and reputation systems~\cite{Mos04}. In these we no longer consider the aggregation of preferences into a global/shared ranking, but instead seek to provide personalized rankings or recommendations to each agent. 
A fundamental challenge in this context is the search for effective \emph{trust-based recommendation systems}, in which -- based on trust-relationships among the agents, and expressed opinions of a subset of them about a service or a product -- a recommendation about a service or a product is provided to agents who did not evaluate it personally. The puzzling challenge of generating useful trust-based recommendation systems is amenable to an axiomatic treatment, beginning with an attempt to characterize the systems satisfying different sets of desired properties. 

In this paper we significantly expand the body of work on the axiomatic approach for internet settings by initiating work on the axiomatic treatment of {\em Group Recommendation Systems}. We assume a trust-graph as described above, where agents express who they trust, and information is provided about the opinions of some of the agents about a product/service, but we care about providing recommendations to a {\em group} of agents, rather than a single one (e.g., a party of friends, looking for a restaurant). Notice in this case the group may include some agents who have experienced the service/product directly and some who may have not. This can be viewed as a bridge between social choice (aggregating individual preferences) and trust-based recommendation systems. In addition to its theoretical importance, the topic of group recommendation systems is of great practical interest: from companies recommending a new networked game to a group of players just finishing one, through recommending a TV show to watch as a family\cite{ZGMZG15}, to a group of friends looking for a holiday destination.

While group trust-based recommendation systems are vastly different from (individual) trust-based recommendation systems, for reasons of exposition we will do our best to connect to the literature on the latter, adopting (and adapting) properties taken as axioms for individual trust-based recommendation systems in earlier work, and adding to them properties reflecting the fact we deal with group recommendations. 
Interestingly, putting these together lead to a powerful and illuminating impossibility result; the axioms/properties are all essential for this impossibility, as removing each one of them leads to possibility. Given this impossibility, we replace our three group related axioms by three other properties; in a second major result we show an extension of a random walk system satisfying all desired properties, and moreover it is the {\em only} system satisfying these properties. Together, this provides rigorous foundations to a theory of group recommendations. 

Following the basic model definitions, we present an overview of the axioms considered and their motivations. We re-emphasize that these axioms are either properties accepted in the foundational work on individual trust-based recommendation systems, or minimal properties capturing our aim at \emph{group} recommendations. Then we present the general impossibility theorem. Finally, re-visiting the group recommendations' axioms, we show a full unique characterization of a group recommendations random walk system.

\section{Related Work}
Most approaches to individual recommendation systems~\cite{RV97} (i.e., systems not attempting to recommend to groups) have proposed a model based on their observations of recommendation dynamics in real life, without setting out to achieve any particular mechanism behavior. Much work is devoted to collaborative filtering, in which an agent receives recommendations based on the views of agents with similar properties -- similar to the design of the Netflix challenge~\cite{BL07,GS09}. Other work focuses on simulations and field experiments~\cite{SKR99,RZ02,AMT05}. Some models add a social graph to the recommendation system, supporting a different trust level for each agent. However, these works \cite{WBS08,PMLR04,HS10} use the social graph as a mechanism which propagates true, objective information to the agents, and do not consider agents' recommendations as opinions which may depend on taste (and hence, have no fixed value of ``trustworthiness''). 

In the past few years, more research has been devoted to group recommendations, as the scenarios where group recommendation are useful are more and more evident. Early work has simply aggregated all members' preferences~\cite{OCKR01,GSOG09}, but the common approach tries to implement a model which adds a layer of complexity beyond agents' approval of a choice, emphasizing the importance of the group itself \cite{YCL14,GLRW13}, like adding also the measure of the rejection of a choice by each participant \cite{ABCDY09,KKOR10,GXLBHMS10} or trying to build a power relationship between participants~\cite{SYMM11}. Gartrell et al. \cite{GXLBHMS10} try using the social graph, but ultimately the approach is limited as it is used only to propagate information, and not in the actual recommendation system.

A different approach to recommendation systems is the axiomatic one, which seeks to first describe the goals of a system, and then to find the systems that implement such goals. Such an approach has been taken in ranking systems \cite{AT05,AT10}, including detailed analysis of specific mechanisms~\cite{AT05b} including collaborative filtering~\cite{PHG00}. More importantly, it has been applied to the individual recommendation system problem, first in Andersen et al.\cite{ABCFFKMT08}, and following that, in additional papers complementing it~\cite{RT09,BCTMT10}. One of the key strengths of this research path is in its basic model, which incorporates the influence of the social graph on agents' behavior. Alas, these papers do not deal with group recommendations, and hence with the particular needs and desired properties of this problem.

\section{Preliminaries and Model Definitions}

The basic model (adapted from~\cite{ABCFFKMT08,RT09,BCTMT10}), deals with a graph that has opinionated nodes (or voters) over some option -- some are $+$ nodes (agents which like the option), and some are $-$ nodes (the agents which did not like the option). The rest of the nodes are nonvoters, i.e., they have no predetermined opinion. A directed edge $(a,b)$ indicates that the agent $b$ influences agent $a$'s opinion to some degree. We wish to find a mechanism that takes any group of agents (both voters and nonvoters) and gives the members of the group a single recommendation -- $+$, $-$, or $0$ (in case of inability to recommend). Formally:

\begin{definition}
A {\bf voting network} is a directed graph $G(N,V_{+},V_{-},E)$, where $N$ are the nodes, $V_{+}\subseteq N$ are the nodes which vote $+$, $V_{-}\subseteq N$ are nodes which vote $-$, and $E$ are directed edges, in which parallel edges are allowed\footnote{Equivalent to using weights, but easier to analyze in our case. This means $E$ is, in effect, a multiset.}, but not self loops. We say node $b\in N$ influences node $a\in N$ when there exists an edge $(a,b)$.
\end{definition}

From this definition we can derive the group of voters -- $V_{+}\cup V_{-}$ and nonvoters -- $N\setminus(V_{+}\cup V_{-})$.

\begin{definition}
A {\bf group recommendation system} is a function $R_{G}:2^{N}\rightarrow \{+,-,0\}$, assigning a recommendation to each subset of graph nodes in the graph $G$.
\end{definition}

Before proceeding to the axioms' definitions, we define a group random-walk recommendation system variant. For this we first define an individual random walk recommendation system, which, basically, assigns to each node the sign of the weighted average of all voters which are reachable from it.

\begin{definition}
An {\bf individual random walk recommendation system} takes a voting network\linebreak$G(N,V_{+},V_{-},E)$ and assigns each node $a\in N$ a value $r_{a}$: If $a\in V_{+}$ (respectively, $a\in V_{-}$), then $r_{a}=1$ (respectively, $-1$). If $a$ is a nonvoter which does not have a path to any voter, $r_{a}=0$. If it does have paths to other voters, we look at the group $succ_{a}=\{b| (a,b)\in E\}$, and define $r_{a}=\frac{\sum_{b\in succ_{a}}r_{b}}{|succ_{a}|}$. Once calculated $r_{a}$ (based on other $r_{i}$s), the recommendation system recommends $sgn(r_{a})$ (i.e., sign of the number).\footnote{That this is, indeed, a random walk, and $r_{a}$ is equal to the probability of reaching a vertex in $V_{+}$ minus the probability of reaching a vertex in $V_{-}$, and that it is unique is shown in Section 3.1 of \cite{ABCFFKMT08}.}
\end{definition}

Group random-walk, in a sense, calculates a value for each group member based on their individual recommendations, and sums over all of the group members.

\begin{definition}
A {\bf group random walk recommendation system} takes a group $C\subseteq N$ and for each $c\in C$ assigns $r_{c}$ to be the value (not just the sign) of the same node under the individual recommendation system (so $-1\leq r_{c}\leq 1$). It then returns $sgn(\sum_{c\in C}sgn(r_{c}))$.
\end{definition}

\section{The Axioms and their Motivation}

The axioms and their formulation are key to this paper, and therefore we expand on their motivation and intuitive understanding. Formal definitions for each axiom appear following this explanation.

\subsection{The Basic Axioms}
These axioms were adapted from the individual recommendation case, and are quite basic, so that we believe most general-use systems which rely on the social graph for their recommendations would seek to implement them:

\begin{enumerate}

\item {\bf Anonymity} -- \emph{No node is special.} Isomorphic graphs (including recommendation isomorphism -- $+$ and $-$ vote symmetry) have isomorphic recommendations.
\end{enumerate} 
\begin{axiom}
{\bf Anonymity:} Let $G(N,V_{+},V_{-},E)$ be a voting network, and $R$ a recommendation system. For any permutation $\pi:N\rightarrow N$ and $G'$, the isomorphic voting network under it, for any $C\subseteq N$ $R_{G}(C)=R_{G'}(\pi(C))$. Furthermore, For $G''(N,V_{-},V_{+},E)$ and $C\subseteq N$, $R_{G}(C)=-R_{G''}(C)$.
\end{axiom}

It is natural that if a group gets a certain recommendation (w.l.o.g, $+$), and a $+$ voter joins the group or gets additional influence over it, the recommendation should not change.

\begin{enumerate}
 \setcounter{enumi}{1}

\item {\bf Positive Response} -- \emph{Adding support for a recommendation cannot reverse it.} A group recommended $+$ to which a $+$ voter is added (or begins to influence its members) does not change its recommendation. If a group is recommended 0, adding a $+$ voter to the group changes the vote to $+$. Furthermore, adding an unconnected $+$ voter and a $-$ voter (both not in the group), both influencing the same node in the group does not change the recommendation.

\end{enumerate}
\begin{axiom}
{\bf Positive Response:} Let $G(N,V_{+},V_{-},E)$ be a voting network, $R$ a recommendation system, $C\subset N$, and $a\in V_{+}\cap\{N\setminus C\}$ such that there is no edge $(c,a)\in E$ for any $c\in C$ . Then if $R_{G}(C)=+$ or $R_{G}(C)=0$, then both $R_{G}(C\cup \{a\})=+$, and if we define $G'$ as $G$ with an added edge $(c,a)$ for some $c\in C$, $R_{G'}(C)=+$.

Furthermore, let $b\in V_{-}\cap\{N\setminus C\}$ such that there is no edge $(v,b)\in E$ for any $v\in N$ nor edges $(v,a)\in E$ for any $v\in N$ (i.e., $a$ and $b$ are isolated). For any $c\in C$ define $G''(N,V_{+}, V_{-}, E\cup \{(c,a), (c,b)\})$ then $R_{G}(C)=R_{G''}(C)$.
\end{axiom}

As we are investigating members of a social graph, it makes sense to ignore nodes that are not in the same connectivity group as members of the group, hence:
\begin{enumerate}
 \setcounter{enumi}{2}
\item {\bf Independence of Irrelevant Stuff (IIS)} -- \emph{Unrelated nodes do not affect recommendation.} A node's recommendation is only dependent on the nodes that it can reach. Voters are not, of course, influenced by any edges, as their opinion is already set, so removing their outgoing edges has no effect.
\end{enumerate}

\begin{axiom}
{\bf IIS:} Let $G(N,V_{+},V_{-},E)$ be a voting network, $R$ a recommendation system, and $C\subset N$. If $d\in N$ is not reachable from $C$, then let $E_{d}\subseteq E$ be the set of edges outgoing or incoming from $d$, and define $G'(N\setminus d, V_{+}\setminus d, V_{-}\setminus d, E\setminus E_{d})$, then $R_{G}(C)=R_{G'}(C)$. Furthermore, if $e\in E$ an edge $(a,b)$ for a voter $a$ (i.e., $a\in V_{+}\cup V_{-}$ and is being influenced by $b$), then for $G''(N,V_{+},V_{-},E\setminus \{e\})$, for every $C\subset N$, $R_{G}(C)=R_{G''}(C)$.
\end{axiom}

\subsection{Group Power Axioms}

We now turn to axioms which try to portray the unique properties of a group recommendation system. In such systems, we want the group members to have a larger influence on the decision than external agents, though we do wish to allow external influence in some cases. In order to portray these two, somewhat conflicting, desires, we define very limited, narrow, axioms, only on a particular structure of social graph -- \emph{star groups}, which (as shown in Figure~\ref{alphaEx}), are made of a certain type of voters in the group (w.l.o.g, $+$), which influence all of the group's nonvoters, with $-$ voters influencing these nonvoters from outside the group.

\begin{definition}
A group $C$ with $n$ voters $\{v_1,\ldots ,v_n\}$ and $m$ nonvoters $\{u_1,\ldots ,u_m\}$ is a {\bf star group} (e.g., Figure~\ref{alphaEx}) if:
\commentout{
(1) All voters $v_j\in C$ have the same label (w.l.o.g., $+$). (2) For every nonvoter $u_i\in C$ and voter $v_j\in C$ there exist an edge $(u_i,v_j)$ (so each nonvoter is connected to all voters in the group). (3) Every nonvoter $u_i\in C$ has an associated group, $D^i=\{t|$ there is an edge $(u_i,t)$ for voter $t\notin C\}$. For every $i$, all $D^i$ members are labeled $-$, and for every $i,h$: $D^i\cap D^h=\emptyset$. (4) Nonvoters $u_{i}\in C$ have no other edges.
}
\begin{itemize}
\item All voters $v_j\in C$ have the same label (w.l.o.g., $+$).
\item For every nonvoter $u_i\in C$ and voter $v_j\in C$ there exist an edge $(u_i,v_j)$ (so each nonvoter is connected to all voters in the group).
\item Every nonvoter $u_i\in C$ has an associated group, $D^i=\{t|$ there is an edge $(u_i,t)$ for voter $t\notin C\}$. For every $i$, all $D^i$ members are labeled $-$, and for every $i,h$: $D^i\cap D^h=\emptyset$,
\item Nonvoters $u_{i}\in C$ have no other edges.
\end{itemize}
\end{definition}

%
Having defined star-groups, we now define a set of axioms which are narrow in scope -- only applying to star-groups:

\begin{enumerate}
 \setcounter{enumi}{3}
\item {\bf $\alpha$-\emph{centripetal}} -- \emph{Members of a group have more influence over the recommendation than voters outside it.} If the star-group has $k$ $+$ voters, the recommendation will be $+$ as long as each nonvoter is connected to less than $\alpha k$ $-$ voters (see Figure~\ref{alphaEx}).
\end{enumerate}

\begin{axiom}
{\bf $\alpha$-\emph{centripetal}}: A recommendation system has some $\alpha \in \mathbb{R_{+}}$, $\alpha\geq 1$, such that for every star group (whose members vote, w.l.o.g., $+$) for which for every $i$, $|D^i|\leq \alpha\cdot n$, the recommendation for the group is $+$.
\end{axiom}

However, we do not want the group to be all powerful. When there are few $+$ voters in the star-group, and many nonvoters and $-$ voters, we would like to see some influence of the outside agents. Thus:

\begin{enumerate}
 \setcounter{enumi}{4}
\item {\bf $(\beta,r)$-\emph{centrifugal}} -- \emph{Agents outside the group may still influence it.} If a star-group (whose members vote, w.l.o.g., $+$) has $k$ $+$ voters, but even more nonvoters -- more than $r$ nonvoters for each $+$ voter; and each nonvoter is connected to many $-$ voters -- at least $\beta k$ -- the group recommendation would be $-$ (see Figure~\ref{betaEx}).
\end{enumerate}

\begin{axiom}
{\bf $(\beta,r)$-\emph{centrifugal}}: A recommendation system has some $\beta \in \mathbb{R_{+}}$, $\beta\geq 1$ such that for every star group for which $\frac{m}{n}\geq r$ ($r\in\mathbb{R}_{+}$) and for which for every $i$, $|D^i|\geq \beta\cdot n$, the group's recommendation is $-$.
\end{axiom}


\begin{example}
To illustrate the two axioms above, we first reiterate that they are extremely narrow, i.e., they do not apply to almost any graph, only to those constructed as a star group (as in Figure~\ref{alphaEx}). Other groups, whatever their makeup (e.g., the sub-groups in Figure~\ref{dividedState}, which are connected to a variety of $+$ and $-$ nodes), are not impacted at all by these axioms, and any recommendation system is free recomend as it sees fit in these cases. Moreover, some star groups do not fall under the purview of these axioms either, if they do not fully comply with their conditions.

\begin{figure}
\begin{center}
\includegraphics[scale=0.4]{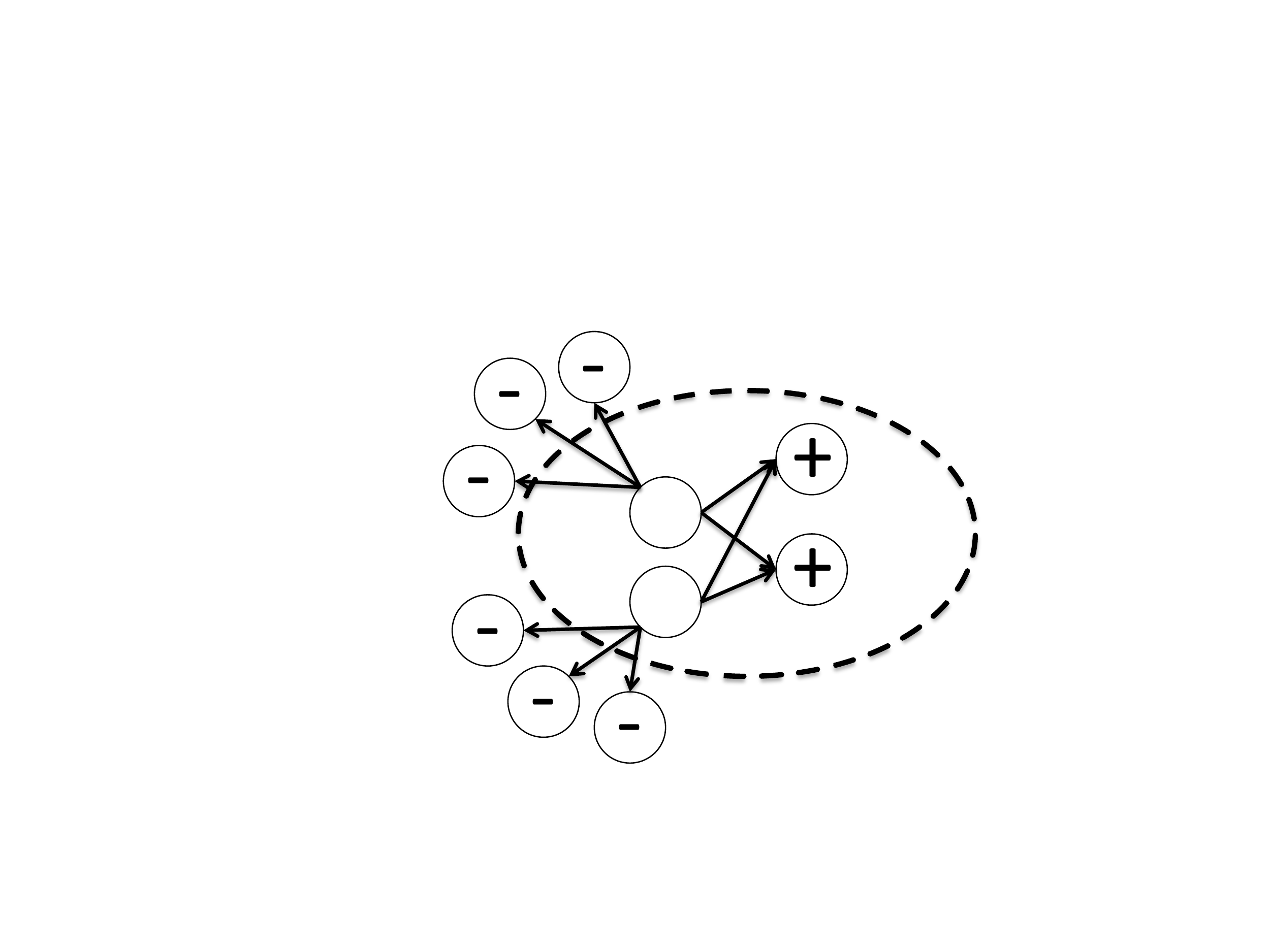}
 \end{center}
\caption {For $\alpha>1\frac{1}{2}$, the $\alpha$-centripetal axiom on this star group implies the recommendation here will be $+$.}\label{alphaEx}
\end{figure}

Illustrating $\alpha$-\emph{centripetal}, examine Figure~\ref{alphaEx}. For $\alpha>1.5$, the recommendation for this group is $+$. This is since for each of the nonvoters, the ``influence'' of the $+$ nodes, which are in their group, is greater than that of the $-$ nodes, which are outside it. The $\alpha$-\emph{centripetal} axiom tries to capture this greater influence of the the nodes inside the group over the ultimate recommendation for the group.


\begin{figure*}[t!]
    \centering
    \begin{subfigure}[t]{0.5\textwidth}
        \centering
        \includegraphics[scale=0.5]{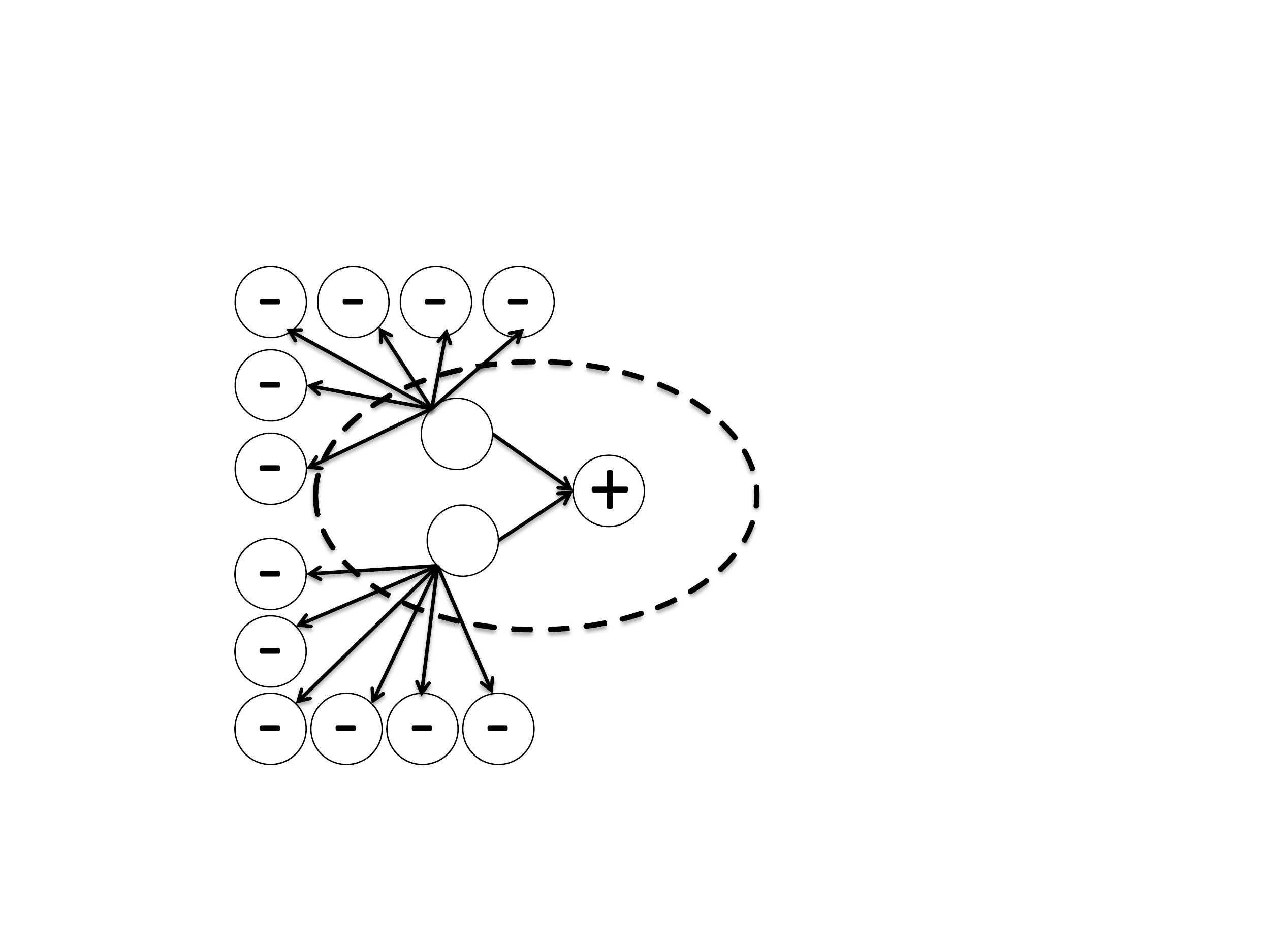}
  \caption{For $\beta\leq 6$, $r<\frac{2}{3}$, the $(\beta,r)$-\emph{centrifugal} axiom on this star group implies the recommendation here will be $-$.}\label{betaEx}
      \end{subfigure}%
    ~ 
    \begin{subfigure}[t]{0.5\textwidth}
        \centering
        \includegraphics[scale=0.3]{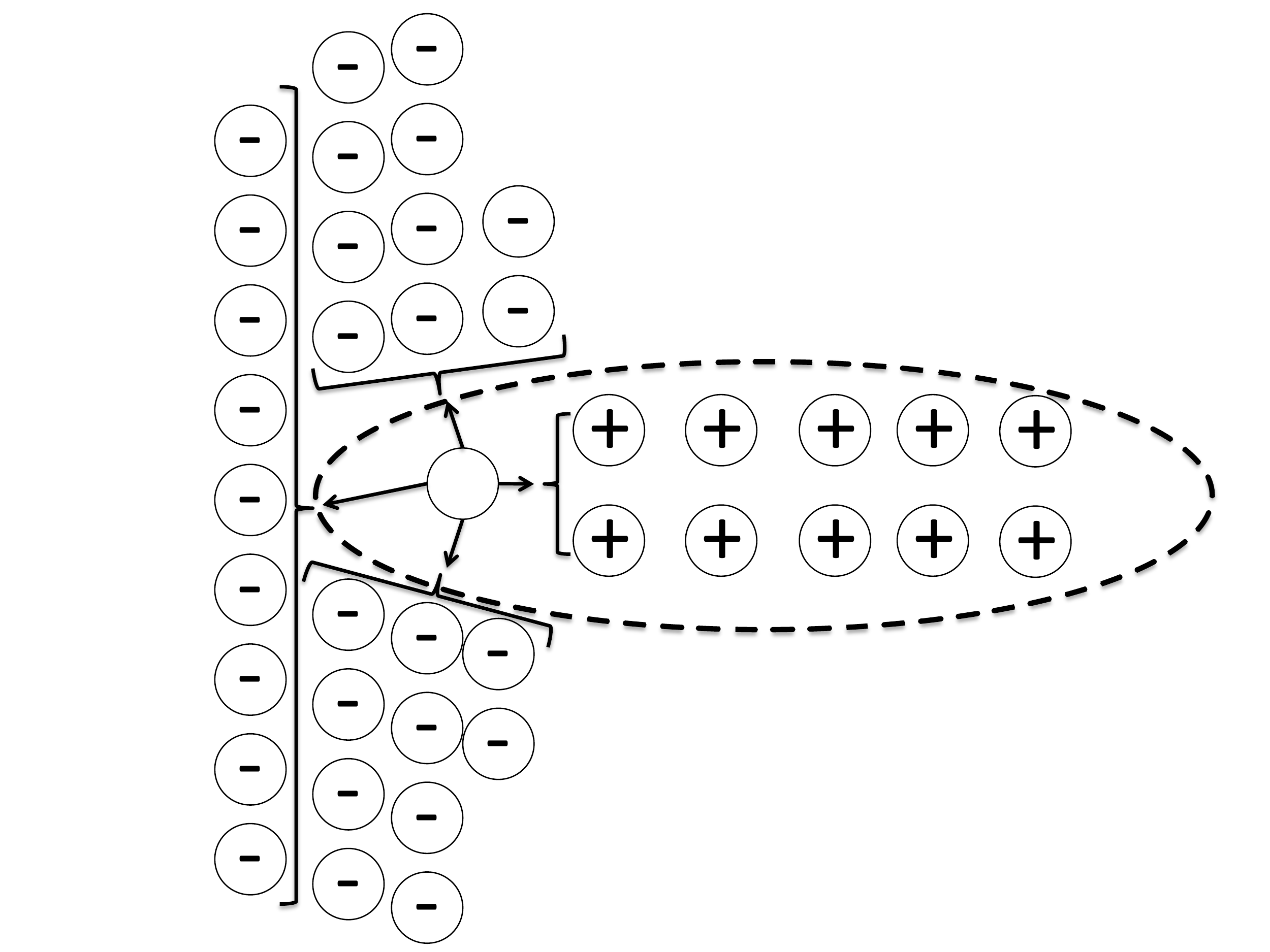}
\caption{For $r>\frac{1}{11}$, the $(\beta,r)$-\emph{centrifugal} axiom cannot be applied to this star group, regardless of the value of $\beta$.}\label{rEx}
    \end{subfigure}
    \caption{$(\beta,r)$-\emph{centrifugal} examples}
\end{figure*}

Now, taking a look at $(\beta,r)$-\emph{centrifugal} axiom. Let us first focus on the $\beta$ part. Its role is to ensure that if there is enough external influence, the external nodes will have some effect (otherwise, external nodes could just be ignored). I.e., there is some (large) number, such that if enough $-$ nodes are connected to the nonvoters, they will sway the group towards $-$. In cases such as that in Figure~\ref{betaEx}, for any $\beta\leq 6$, there are more than $\beta$ $-$ nodes influencing each undecided node in the group vs. only a single $+$ node in the group, so the recommendation would be $-$.
%

Now turning look at $r$. This parameter is used to prevent cases as in Figure~\ref{rEx}, where even with a sufficient $\beta$ (suppose in this case, $\beta\leq 3$), there are many $+$ nodes in the group, far overwhelming the number of nonvoters, so even if these undecided are heavily influenced towards some decision, it should not have any effect on the group's recommendation. If $r>\frac{1}{10}$, the $(\beta,r)$-\emph{centrifugal} axiom is not applicable to Figure~\ref{rEx}, as the ratio of nonvoters to voters in the group is $\frac{1}{11}$ (so any recommendation system may choose to do whatever it wants in this case). 
\end{example}

\begin{enumerate}
 \setcounter{enumi}{5}
\item {\bf Internal consistency} -- \emph{If all of a group's partitions have the same recommendation, that will be recommendation of the whole.} If all sub-groups in a disjoint partition of a group of agents, are given the same (non-neutral) recommendation (and there are no contradicting unanimous, non-neutral partitions), the whole group will have this recommendation as well.
\end{enumerate} 

\begin{axiom}
{\bf Internal consistency}: In a recommendation system $R$, for every $C\subseteq N$, for some partition $C=C_1\cupdot C_2\cupdot\ldots\cupdot C_n$ for which $R_{G}(C_1)=R_{G}(C_2)=\ldots=R_{G}(C_n)\neq 0$, and if all other similar partitions $C=C'_1\cupdot C'_2\cupdot\ldots\cupdot C'_n$, for which $R_{G}(C'_1)=R_{G}(C'_2)=\ldots=R_{G}(C'_n)\neq 0$ have $R_{G}(C_{1})=R_{G}(C'_{1})$, then $R_{G}(C)=R_{G}(C_1)$.
\end{axiom}

Ultimately, we will show the above six axioms are incompatible, and there exist no group recommendation system that can accommodate them. We will now consider three additional axioms. 

\subsection{Influence Structure Axioms}

The final three axioms, are, to a certain extent, a group-recommendation extension of axioms suggested for the individual recommendation case \cite{ABCFFKMT08}. They capture influence and the way it ``moves'' through the social connections, such that influence can extend beyond an immediate node (so one may influence a second person, and that person may, in turn, influence another).

\begin{figure}
\begin{center}
\includegraphics[scale=0.35]{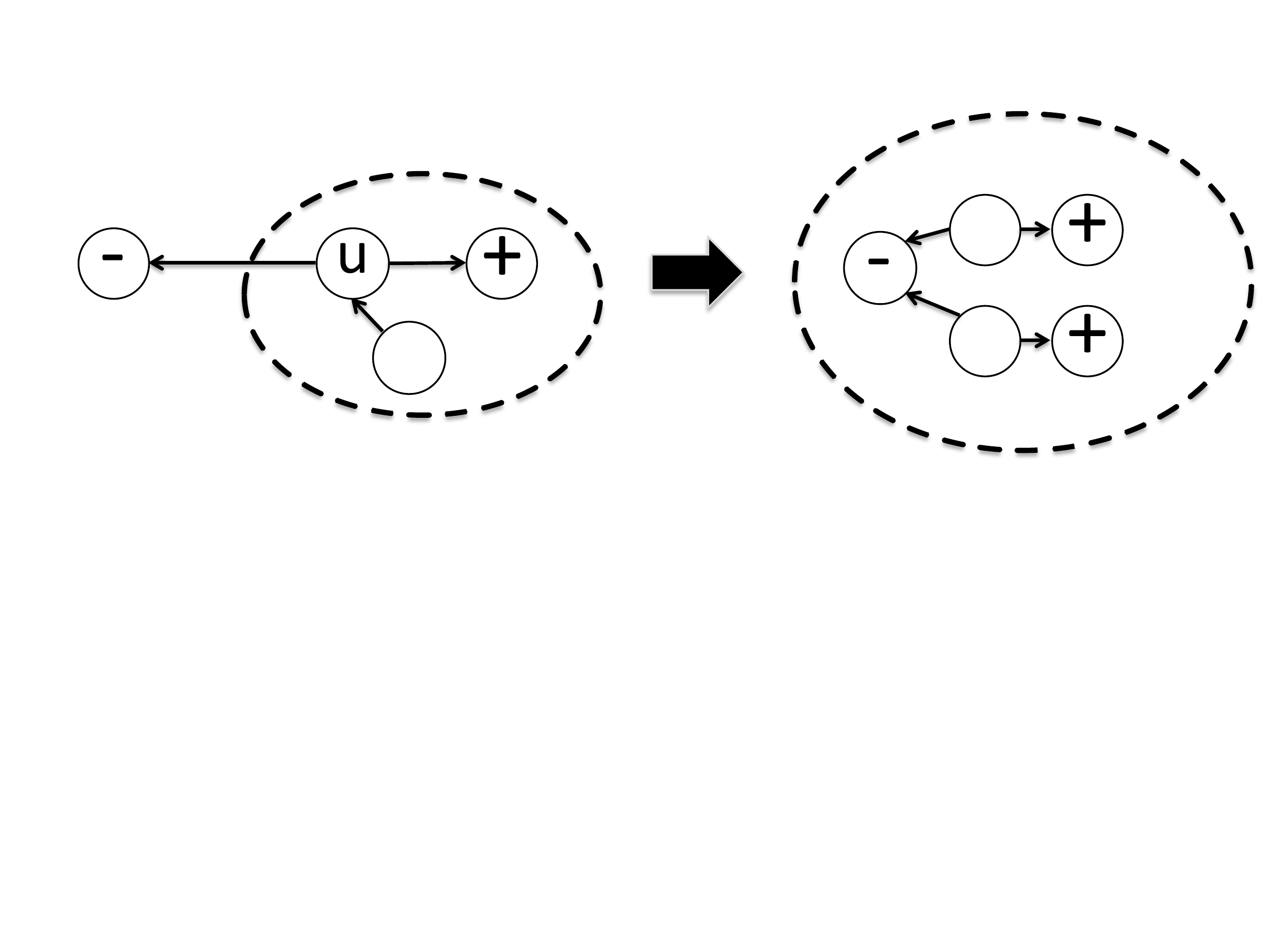}
 \end{center}
\caption {Example of applying proportional inclusiveness on $u$: The $-$ node is now inside the group, but since its weight on $u$'s recommendation was half the total of influences on $u$, the other nodes in the group are duplicated accordingly.
}\label{propDraw}
\end{figure}

\begin{enumerate}
 \setcounter{enumi}{6}
\item {\bf Trust Propagation} -- \emph{Influence moves along the graph.} If nonvoter $b$ has $k$ edges to nodes influencing it, and node $a$ is influenced by $b$ with $k$ edges, then the edges to $b$ can be replaced by edges to the nodes influencing $b$.
\begin{axiom}
 {\bf Trust Propagation}: Consider recommendation system $R$, voting network\linebreak$G(N,V_+,V_-,E)$, group $C\subseteq N$, and nonvoters $u,v\in N$ for which the edges leaving $v$ (beside $(u,v)$) are $(v,w_1)\ldots (v,w_k)$ for some $k\geq 1$. Suppose $E$ contains $k$ copies of $(u,v)$, and we construct $E'=(E\cup \{(u,w_1),\ldots (u,w_k)\}\setminus \{(u,v)\cdot k\})$ and $G'(N,V_+,V_-,E')$, then $R_{G}(C)=R_{G'}(C)$.
\end{axiom}

\item {\bf Scale Invariance} -- \emph{Influence does not care about units.} Duplicating a node's outgoing edges (i.e., edges to the nodes influencing it) does not change a recommendation.

\begin{axiom}
{\bf Scale Invariance}: For a voting network $G(N,V_+,V_-,E)$, and a nonvoter $u$, the recommendations are identical for $G'(N,V_+,V_-,E\cup E')$ where $E'$ contains $k$ copies of each of $u$'s outgoing edges.
\end{axiom}

\item {\bf Proportional Inclusiveness} -- \emph{An external influence can be described as a group-member influence.} A voter outside a group, connected directly to a nonvoter inside it, has an influence over the group recommendation in proportion to its weight of influence on the nonvoter, and the nonvoter's influence in the group. Therefore, the recommendation for a group would be the same as for a group that includes also the voters influencing a nonvoter in the original group (with a few adjustments to maintain relative power of group members, see Figure~\ref{propDraw}).
\begin{axiom}
{\bf Proportional Inclusiveness}: For a voting network $G(N,V_{+},V_{-},E)$, a group $C\subseteq N$, a nonvoter $u\in C$ and voters $v_{1},\ldots, v_{m}\in V\setminus C$ and $v_{m+1},\ldots,v_{t}\in C$ which are influencing it (i.e., $(u,v_{i})\in E$) then the following transformation retains recommendations:
Let there be $k_{i}$ copies of $(u,v_{i})$ in $E$, and $s$ edges $(u,*)$ in $E$ (i.e., $s=\sum_{i=1}^{t}k_{i}$). For $1\leq j\leq s$ we define $N^{j}=N\setminus\{u\}$ and define $N'=\cup_{j=1}^{s}N^{j}\cup \{v_{m+1}\}^{k_{m+1}}\cup\ldots\cup \{v_{t}\}^{k_{t}}$. For each $1\leq i\leq m$ we choose $k_{i}$ nodes of type $v_{i}$ (there are $s$ copies of these in $N'$), and mark them $v_{i}^{1},\ldots,v_{i}^{k_{i}}$.
For each $N^{j}$ we define $C^{j}=C\setminus \{u\}$ and $E^{j}=E\setminus\{(*,u), (u,*)\}$ (no edges ingoing or outgoing from $u$), and tweak it a little: For each $c\in C$ such that there is an edge $(c,u)\in E$, we multiply $s$ times each edge $(c,*)\in E^{j}$, and add $k_{i}$ edges $(c,v_{i})$ for $1\leq i\leq t$ (excluding self, of course). We define $E'=\cup_{j=1}^{s}E^{j}$, and $C'=\cup_{j=1}^{s}C^{i}\cup_{h=1}^{m}\cup_{r=1}^{k_{h}}v_{h}^{r}$.

Now, for $G'(N',V'_{+},V'_{-},E')$, $R_{G}(C)=R_{G'}(C)$.\\
\end{axiom}
\end{enumerate}

These axioms will prove to uniquely characterize a group variant of the random walk recommendation algorithm.

\begin{proposition}\label{indyProp}
Axioms $1-6$ and axioms $1-3,7-9$ (the sets we deal with) are all independent of one another. 
\end{proposition}
\begin{proof}
We list some odd mechanisms that are consistent with each 5 of our axioms, demonstrating the necessity of each (we present here the main points of the odd mechanism -- the complete version can be constructed by using the other axioms). First, we begin with axioms 1--6.
\\
\begin{itemize}

\item {\bf Anonymity:} A mechanism that recommends $+$ for every singleton.

\item {\bf Positive response:} A system for which if the group contains a nonvoter which is influenced by both $+$ and $-$ nodes from outside the group, it is recommended 0.

\item {\bf IIS:} Except for star groups, all nodes outside a group are considered to be influencing all nodes inside it.

\item {\bf $\alpha$-\emph{centripetal}:} All star groups are always recommended the opposite of the nodes inside the group (so recommended $-$ even in star groups consisting of arbitrarily large $M$ of $+$ voters, and one nonvoter connected to a sole $-$ voter).

\item {\bf $(\beta,r)$-\emph{centrifugal}:} All star groups are always recommended as their internal nodes (so recommended $+$ even in star groups consisting of one $+$ voter, arbitrarily large $M_{1}$ of nonvoters, each connected to arbitrarily large $M_{2}$ $-$ voters.

\item {\bf Internal consistency:} Taking 3 nonvoters, 3 $+$ voters, and a single $-$ voter. Connecting each nonvoter to a single $+$ voter, and two of the nonvoters are also connected to the $-$ voter (see Figure~\ref{interExample}). Our groups are made of the pairs of $+$ voters with the nonvoters connected to them. When we take a single pair, one that is also connected to the $-$ voter, the recommendation is $+$. Adding another pair that is not connected to the $-$ voter, the recommendation is now $-$.

\begin{figure}
\begin{center}
\includegraphics[scale=0.5]{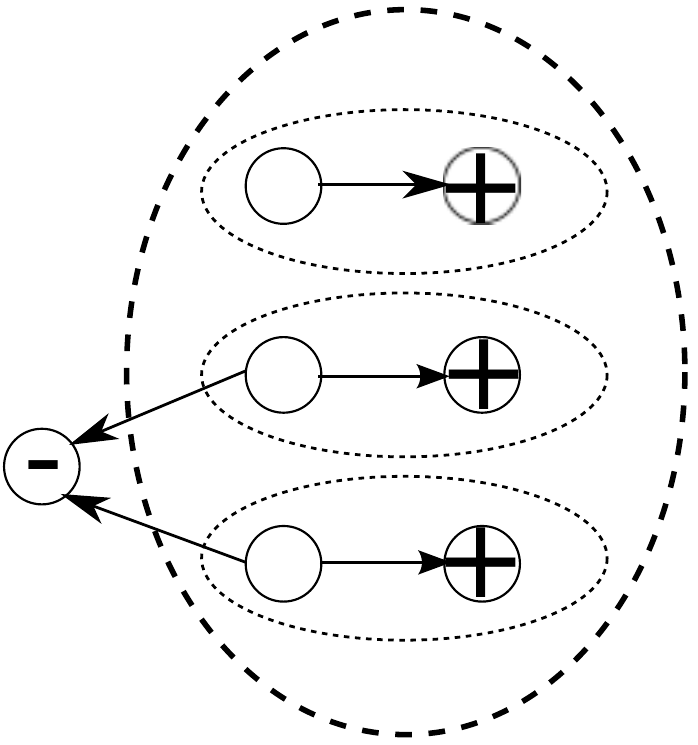}
 \end{center}
\caption {Internal consistency problem}\label{interExample}
\end{figure}

\end{itemize}

We now continue with axioms 1--3 and 7--9: 

All examples given in appendix 1 of \cite{ABCFFKMT08} for axioms 1--3 work. Hence, we just need to show for 7--9:

\begin{itemize}

\item {\bf Trust propagation:} A group is recommended by summing for each vertex over number of $+$ and $-$ nodes it is directly connected to.

\item {\bf Scale invariance:} A group is recommended by a random walk from each vertex in the group, but to the relative influence of each edge is added the number of outgoing edges from the influencing vertex.

\item {\bf Proportional inclusiveness:} A group ignores all outside influences -- performs simple majority on $+$ and $-$ nodes inside it.
\end{itemize}
\end{proof}
\section{An Impossiblity Result}\label{impossibleSec}

When looking at the axioms detailed above, it is clear that axioms $1-3$ are imperative for any basic social graph based recommendation system, whether it is an individual or a group one. However, when expanding to group recommendation, one appreciates the goal that the nodes inside a group will have some different -- stronger -- effect than those outside it, but that this effect will not be unbounded, so outside information will not be completely ignored. Axioms $4$ and $5$ distill this goal in a most clear-cut way (on \emph{star groups}), being intentionally narrow and non-sweeping. To that we add axiom $6$, which is an intuitive and desirable consistency requirement.

However, these requirements are not compatible:

\begin{theorem}\label{impossibleTrm}
No recommendation system satisfies axioms $1-6$, i.e., is anonymous, positive responsive, IIS, is internally consistent and is $\alpha$-\emph{centripetal} and $(\beta,r)$-\emph{centrifugal} for $1<\alpha,\beta<\infty$, $0<r<\infty$.
\end{theorem}
\begin{proof}
First, we note that if the $(\beta,r)$-\emph{centrifugal} axiom uses an $r\notin\mathbb{N}$, then we shall use as $r:=\lceil r\rceil$.

Our proof is constructed using 3 steps in which we build a graph, find a specific group in it, and show that the axioms require it to both be recommended $+$ and $-$, creating a contradiction.
\\

{\bf Step 1: Build a graph}

Since $\alpha>1$, there is some $\ell\in\mathbb{N}$ for which $\ell\alpha-(\ell+2)\geq 0$. We define $k$ as $\lceil\frac{\beta-1}{\alpha-1}\rceil + \ell$ and $s$ as $\lfloor k\alpha\rfloor$. We now build the following graph, consisting of a star group containing $k$ $+$ voters $(v_1,\ldots ,v_k)$ and $k\cdot r$ nonvoters $(u_1,\ldots, u_{kr})$, and outside the star group are $s\cdot k\cdot r$ $-$ voters $(t_1,\ldots, t_{skr})$, with each nonvoter connected to $s\cdot r$ $-$ voters (i.e., every nonvoter $u_i$ has the edges $(u_i,t_h)$ for $h\in\mathbb{N}$, $i\leq h\leq (i-1)+s$).
\\

{\bf Step 2: Build an indivisible positive set}

We now wish to construct a set $C$ which has no partition for which each part is recommended $+$. We call our star group $\tilde{C}$, (with $k$ $+$ voters and $kr$ nonvoters). According to the $\alpha$-\emph{centripetal} axiom, since $s\leq k\cdot \alpha$ for each nonvoter, $\tilde{C}$'s recommendation is $+$. We now seek to find the minimal $+$ part of $\tilde{C}$ -- suppose $\tilde{C}$ has a partition for which every part is recommended $+$. In at least one of these parts the number of nonvoters exceeds (or is equal to) $r$ times its number of voters, and we call it $\tilde{C}'$, and continue the process. This process ends with a set $C$ with $a$ nonvoters and $b$ $+$ voters ($b\leq rb\leq a\leq rk$), for which the recommendation system recommends $+$. We shall now show that according to the $(\beta,r)$-\emph{centrifugal} axiom and internal consistency, it needs to be recommended $-$, causing a contradiction.
\\

{\bf Step 3: Build a contradictory partition}

We need to show a partition which results in every part being recommended $-$, and show that there is no partition for which each part get recommended $+$. The latter is trivial thanks to the previous minimization process -- if there is such a partition, then the process has not ended yet. However, we can partition $C$ into a sets of one $+$ voter and $r$ nonvoters (possibly, some nonvoters end without any voter to group them with -- they are grouped apart). We shall now show that these sets need to be recommended $-$, causing the contradiction.

Since $+$ voters in the set are more heavily weighted that outside it, we can focus our proof just for sets with $r$ nonvoters and one voter, and that will suffice for the case of a lone nonvoter. Note that in these sets our nonvoter is connected to one $+$ voter in its set, $k-1$ $+$ and $s$ $-$ voters outside it. According to the positive response axiom (number 2), we can remove one $+$ and one $-$ voters (when both do not belong to the set) from each nonvoter without changing the recommendation. We are left with one $+$ voter connected to the nonvoter in the set, and $s-(k-1)$ $-$ voters connected to the nonvoter. we now wish to prove that $s-k+1\geq\beta$.

Due to $k$'s definition, we know

\begin{equation*}
\begin{split}
\frac{\beta-1+\ell\alpha-\ell}{\alpha-1}\leq k \leq &\frac{\beta-1+(\ell+1)\alpha-(\ell+1)}{\alpha-1}=\\&=\frac{\beta+(\ell+1)\alpha-(\ell+2)}{\alpha-1}
\end{split}
\end{equation*}

Similarly, we know

$$
s\geq \alpha \frac{\beta-1+\ell\alpha-\ell}{\alpha-1}-1=\frac{\beta\alpha+\ell\alpha^{2}-(\ell+2)\alpha +1}{\alpha-1}
$$

Hence:

\begin{equation*}
\begin{split}
&s-k+1\geq\\&\frac{\beta\alpha+\ell\alpha^{2}-(\ell+2)\alpha+1-\beta-(\ell+1)\alpha+(\ell+2)+\alpha-1}{\alpha-1}=\\=&\frac{\beta\alpha+\ell\alpha^{2}-(2\ell+2)\alpha-\beta+\ell+2}{\alpha-1}=\\=&\frac{(\alpha-1)(\beta+\ell\alpha-(\ell+2))}{\alpha-1}=\beta+\ell\alpha-(\ell+2)
\end{split}
\end{equation*}

Thanks to our definition of $\ell$, this means $s-k+1\geq \beta$, hence the set is recommended $-$, reaching a contradiction.
\end{proof}

\begin{example}\label{exampDesc}
\emph{Showcasing the proof's main parts:} Group of friends are constructed as in Figure~\ref{dividedState}: John, Paul, George, Ringo and Yoko want to go to a restaurant together. John and Yoko have been there and were extremely satisfied with it. However, all the rest of Paul, George and Ringo's acquaintances have a very negative view about the place.

Suppose $\alpha=2$, $\beta=2.5$ and $r=1$. According to $\alpha$-centripetality, the recommendation should be to go the restaurant. However, we can subdivide it into 3 groups (shown in Figure~\ref{dividedState}), which -- using positive response (axiom 2) and $(\beta,r)$-centrifugality (axiom 5) -- should each be advised not to go to the restaurant.

\begin{figure}
\begin{center}
\includegraphics[scale=0.4]{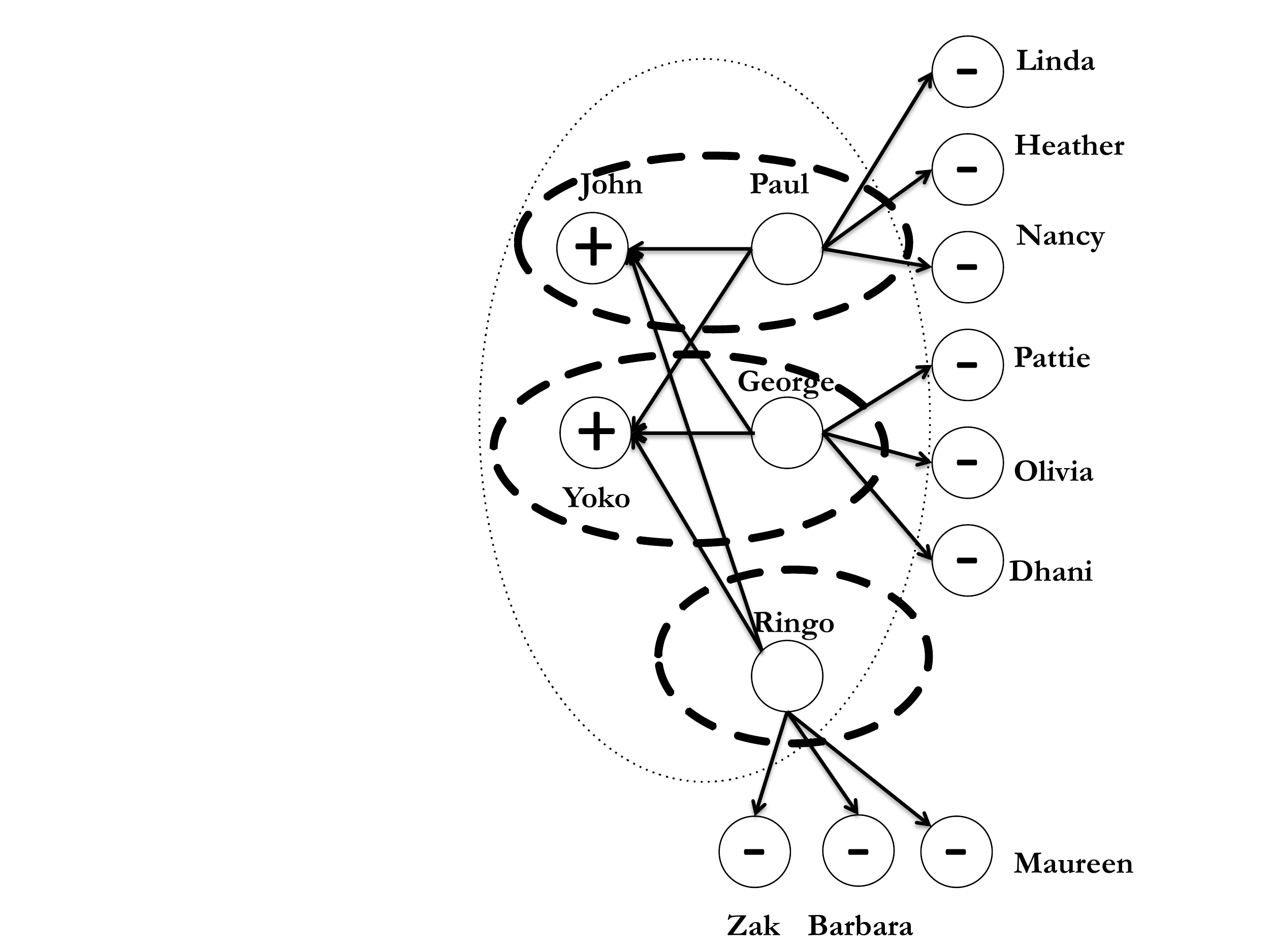}
 \end{center}
\caption {Original group (noted by light line) \& its partition}\label{dividedState}
\end{figure}
\end{example} 

\section{A Recommendation System that Works}

Due to the impossibility result above, we must, of course, give up some of those axioms. We cannot eliminate axioms $1-3$, as they are fundamental to any social graph based recommendation system. However, we replace axioms $4-6$ with axioms $7-9$, which, while they do not allow us the full power of the previous axioms, allow us to show that there is only a single group recommendation system which satisfies them, giving us a potential candidate for a useful, practical system that can be implemented in real-world systems.

\begin{theorem}\label{positiveTheorem}
The group random-walk recommendation system is the only one which satisfies axioms $1-3$ and $7-9$, i.e., is anonymous, positive responsive, IIS, has trust propagation, scale invariance and proportional inclusiveness.
\end{theorem}

\begin{proof}
We shall prove the theorem by looking at a voting system $G(N,V_{+},V_{-},E)$, and looking at a specific group $C\subseteq N$. We will show what the axioms would force its recommendation to be -- and that the same recommendation would be made by the group random-walk recommendation system.
\\

{\bf Part 1: Using the axioms}

Suppose there is an undecided voter $u\in N\setminus C$ which is influenced by $v_{1},\ldots , v_{m}\in N$ (possibly for some $i\neq j$ $v_{i}=v_{j}$) and influences $w_{1},\ldots,w_{t}$ (again, possibly for some $i\neq j$ $w_{i}=w_{j}$). According to scale invariance, recommendations do not change if $w_{i}$ multiplies his connections by $m$. Now, using trust propagation, $w_{i}$ connects directly to $v_{1},\ldots,v_{m}$, and is no longer connected at all to $u$. As we do this for all $w_{1},\ldots,w_{t}$, when we finish, node $u$ no longer influences any other vertex.

Performing these steps for every undecided voter $u\in N\setminus C$ which is influencing and being influenced, we end up with the nodes in the group $C$ either directly connected by an edge -- and being influenced -- to voters or nonvoter sinks (i.e., nonvoters which are not influenced by others). Thanks to the IIS axiom, we can ignore all vertices which are not in the same connected component as $C$. Note that this axiom also means we can ignore all voters or nonvoter sinks which do not directly influence (with a single edge) any node in the group.

Now, using proportional inclusiveness, we eliminate from $C$ all nonvoters which are influenced by voters, leaving in the group, at most, nonvoter sinks (all other members are voters), i.e., there are now $y_{-}$ voters for $-$, $y_{+}$ voters for $+$ and $y_{\circ}$ nonvoter sinks. Suppose $y_{\circ}=0$ -- from anonymity axiom we know if $y_{-}=y_{+}$, recommendation is $0$, and hence from the positive response axiom, the recommendation is $type (\max(y_{-},y_{+}))$ (from the same axiom, that is also the recommendation if $y_{\circ}>0$).
\\

{\bf Part 2: Using the recommendation system}

Now we need to show that the procedure described above reaches the same recommendation as a group random walk would recommend. This recommendation system, in effect, gives all members of $C$ the same weight (say, $1$), and while voters put all their weight on their vote, nonvoters divide their weight according to the random walk. Therefore, we shall show that the contribution of each voter and nonvoter to the final tally is maintained by the changes we do to $C$ in the procedure described above using the axioms.

Furthermore, we note that scale invariance and trust propagation do not affect the result of a random walk, 
hence we can examine the graph as it looks following our multiple applications of these two axioms (just before we begin to apply proportional inclusiveness). Therefore, we need to show that applying proportional inclusiveness does not change the weight in the group. If we manage to show proportional inclusiveness does not change the recommendation of the group random-walk, since the axiom's application leaves us with a group consisting of voters only (and un-influencable nonvoters) we can conduct a simple plurality between the votes, as both group random-walk and the procedure above indicate should happen.

Let the nonvoter we apply proportional inclusiveness to be $u$, which is connected to the voters $v_{1},\ldots, v_{m}\notin C$ and to the nodes $v_{m+1},\ldots, v_{t}\in C$ to each with $k_{i}$ connections (we define $s=\sum_{i=1}^{t}k_{i}$).
There are also the nodes $v_{t+1},\ldots,v_{r}\in C$ which are not connected to $u$. Notice that $u$'s ``vote'' in the group random-walk system gives $\frac{k_{i}}{s}$ weight to $v_{i}$, for $1\leq i\leq t$.

Following an application of proportional inclusiveness, we now have $s$ copies of $v_{t+1},\ldots,v_{r}$ in the new $C$, and $s+k_{i}$ copies of $v_{i}$, $m<i\leq t$. We also have $s$ copies of $v_{i}$, $1\leq i\leq m$, of which $k_{i}$ copies are in $C'$, and we have $s$ copies of any nodes which are not $v_{i}$ ($1\leq i\leq r$), i.e., any node which was not connected to $u$ and not in $C$.

Let us focus on nonvoters in $C\setminus \{u\}$. Each copy of the nonvoter is connected to the same nodes as it was before, except for those which were connected to $u$. Each nonvoter random-walk recommendation before proportional inclusiveness was giving equal weight to each of its connection, so that if a nonvoter had $w$ outgoing edges., hence $\frac{1}{w}$ weight was given to each node connected to it, including $u$. Therefore, $\frac{k_{i}}{s}\frac{1}{w}$ weight for each $v_{i}$ $1\leq i\leq t$ (the nodes connected to $u$). Following proportional inclusiveness, the nonvoter has $w\cdot s$ outgoing edges, the weight of each node it is connected to is $\frac{s}{ws}=\frac{1}{w}$, except the nodes $v_{i}$ ($1\leq i\leq t$), the weight of which is $\frac{k_{i}}{ws}$. Hence the random-walk recommendation for each nonvoter remains the same.

The recommendation for the group $C$ remains the same -- prior to the proportional inclusiveness each node received a weight of $\frac{1}{|C|}$. Following the application of the axiom, $|C'|=s(|C|-1)+s=s|C|$, and each node $c\in C$ such that $c\neq v_{i}$ ($m+1\leq i\leq t$) now has the $s$ copies, and as each copy has the same recommendation of each single one, it contributes the weight of $\frac{s}{s|C|}=\frac{1}{|C|}$. Each node connected to nonvoter $u$ contributed $\frac{k_{i}}{s}\frac{1}{|C|}$, and now each $v_{i}\notin C$ contributes $\frac{k_{i}}{s|C|}$, as there are $k_{i}$ copies of $v_{i}$ in $C'$ for $1\leq i\leq m$. $v_{i}\in C$ now contribute $\frac{s+k_{i}}{s|C|}$, which is the same of $v_{i}$'s contribution as a node of $C$ and as a component of $u$'s recommendation.

Therefore, each node which in the group random-walk would have an influence over the recommendation maintains that level of influence after applying proportional inclusiveness. Multiple applications of this axiom leaves us with a group consisting of purely voters and nonvoter sinks, which according to the axioms leads to a plurality votes among voters, which is exactly the procedure followed by the group random-walk in this case as well.
\end{proof}

\begin{example}
\emph{Showcasing the proof's main parts:} We alter our example, now discussing Peter, Paul and Mary, which are influenced as in Figure~\ref{updatedState}a.
\\

\begin{figure}
\begin{center}
\includegraphics[scale=0.5]{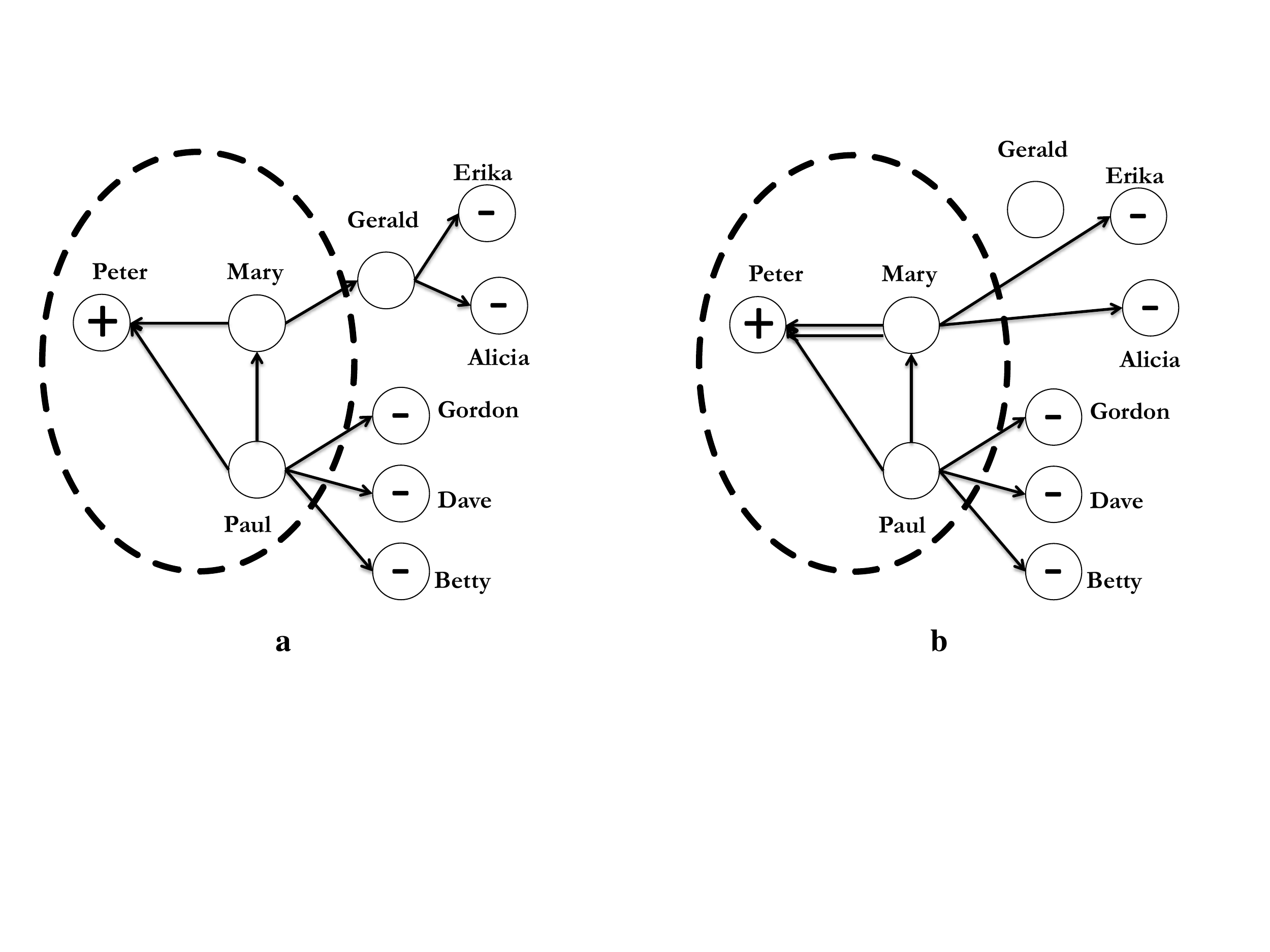}
 \end{center}
\caption {a: Initial state; b: Applying scale invariance and trust propagation }\label{updatedState}
\end{figure}
Using scale invariance and trust propagation, Gerald is no longer connected to Mary. Instead, her attachment to his other influences doubled, and two new connections have been struck (Figure~\ref{updatedState}b). 

%

Using the algorithm for our example means Peter contributes 1, Mary contributes 0, and Paul contributes $-\frac{2}{5}$, meaning the total outcome is still positive, and the restaurant will be recommended.
\end{example}

\section{Discussion and Conclusions}

The challenge of group recommendations has been evident in various online systems as companies struggle to find a way to harness the vast information on people's preferences they accumulate, in order to provide better product recommendations for their users. Most activity in recommendation systems research focus on the individual case, in which a recommendation is given to a single agent. But in many cases, that is not a sufficient response: for example, when a group wishes to decide on a movie or restaurant, when playing a joint game online, coordinating purchases, etc.

The approach proposed in this paper takes a different track than the common one of finding users which resemble those requesting a recommendation and recommending something they found interesting. Rather, we wish to \emph{leverage the social information contained in the social network} -- including both the knowledge of the influences inside a group, as well as the influence of elements outside it on the issues its participants find interest in. This knowledge, we believe, is probably more highly correlated not just with what the group might like, but with what the group would even consider as a relevant option.

In this paper we approached the challenge of group recommendation using the axiomatic approach, enabling certain insight that is not immediately obvious using other methods. We showed that several desirable properties cannot coexist in a group recommendation system, even when we take very narrow, minimalistic, requirements. However, we have been able to show an axiomatization of a variant of a random-walk algorithm that works well on group recommendations. This recommendation system, we believe, is a useful one, and can be used in various real-life settings.

Obviously, we believe future research should focus on developing further methods which utilize the social graph to give more relevant and practical recommendations. This, in our view, is the key to better systems, and ignoring the social context in which groups are formed is to leave out a significant part of the challenge of satisfying users. Naturally, such a progression will include more complex models including, for example, more than 2 possible recommendations.

Furthermore, we believe axiomatization of other recommendation systems may enhance the insight into these methods, allowing better comparisons between the pros and cons of each system. This will enable designers of various online systems to better understand how their choice of mechanism affects the quality of recommendations for the users. Moreover, thinking of additional desired properties for group recommendation systems might lead to new and novel recommendation systems that strive to implement these properties. Our axioms are, in this regard, only a starting point for the community to discuss and consider what are the desirable features of group recommendations.

Finally we note that while some of our desirable axioms were found to be overly restrictive, we believe there still might be a way to include at least part of these axioms in future systems. This will entail replacing one of them -- we tend to believe that internal consistency is the ``easiest'' to let go of -- by other, more lenient ones. This may potentially allow to characterize the family of ``pretty good'' group recommendation systems, providing further foundation for the design and implementation of such mechanisms.

\subsection*{Acknowledgements}
This work was done when both authors where at Microsoft Research, Hertzeliya. The work was also partially supported by NSERC grant 482671.

%
\bibliographystyle{eptcs}
\bibliography{general}  

\begin{thebibliography}{10}
\providecommand{\bibitemdeclare}[2]{}
\providecommand{\surnamestart}{}
\providecommand{\surnameend}{}
\providecommand{\urlprefix}{Available at }
\providecommand{\url}[1]{\texttt{#1}}
\providecommand{\href}[2]{\texttt{#2}}
\providecommand{\urlalt}[2]{\href{#1}{#2}}
\providecommand{\doi}[1]{doi:\urlalt{http://dx.doi.org/#1}{#1}}
\providecommand{\bibinfo}[2]{#2}

\bibitemdeclare{inproceedings}{AT05}
\bibitem{AT05}
\bibinfo{author}{Alon \surnamestart Altman\surnameend} \&
  \bibinfo{author}{Moshe \surnamestart Tennenholtz\surnameend}
  (\bibinfo{year}{2005}): \emph{\bibinfo{title}{On the axiomatic foundations of
  ranking systems}}.
\newblock In: {\sl \bibinfo{booktitle}{Proceedings of the 19th International
  Joint Conference on Artificial Intelligence (IJCAI)}},
  \bibinfo{address}{Edinburgh, Great Britain}, pp. \bibinfo{pages}{917--922}.

\bibitemdeclare{inproceedings}{AT05b}
\bibitem{AT05b}
\bibinfo{author}{Alon \surnamestart Altman\surnameend} \&
  \bibinfo{author}{Moshe \surnamestart Tennenholtz\surnameend}
  (\bibinfo{year}{2005}): \emph{\bibinfo{title}{Ranking systems: The PageRank
  axioms}}.
\newblock In: {\sl \bibinfo{booktitle}{Proceedings of the 6th ACM conference on
  Electronic Commerce (EC)}}, \bibinfo{address}{Vancouver, Canada}, pp.
  \bibinfo{pages}{1--8}, \doi{10.1145/1064009.1064010}.

\bibitemdeclare{article}{AT10}
\bibitem{AT10}
\bibinfo{author}{Alon \surnamestart Altman\surnameend} \&
  \bibinfo{author}{Moshe \surnamestart Tennenholtz\surnameend}
  (\bibinfo{year}{2010}): \emph{\bibinfo{title}{An axiomatic approach to
  personalized ranking systems}}.
\newblock {\sl \bibinfo{journal}{Journal of the ACM}}
  \bibinfo{volume}{57}(\bibinfo{number}{4}), pp. \bibinfo{pages}{1--35},
  \doi{10.1145/1734213.1734220}.

\bibitemdeclare{inproceedings}{ABCDY09}
\bibitem{ABCDY09}
\bibinfo{author}{Sihem \surnamestart Amer-Yahia\surnameend},
  \bibinfo{author}{Senjuti~Basu \surnamestart Roy\surnameend},
  \bibinfo{author}{Ashish \surnamestart Chawlat\surnameend},
  \bibinfo{author}{Gautam \surnamestart Das\surnameend} \&
  \bibinfo{author}{Cong \surnamestart Yu\surnameend} (\bibinfo{year}{2009}):
  \emph{\bibinfo{title}{Group recommendation: semantics and efficiency}}.
\newblock In: {\sl \bibinfo{booktitle}{Proceedings of the VLDB Endowment}},
  \bibinfo{volume}{2}, pp. \bibinfo{pages}{754--765}.

\bibitemdeclare{inproceedings}{ABCFFKMT08}
\bibitem{ABCFFKMT08}
\bibinfo{author}{Reid \surnamestart Andersen\surnameend},
  \bibinfo{author}{Christian \surnamestart Borgs\surnameend},
  \bibinfo{author}{Jennifer \surnamestart Chayes\surnameend},
  \bibinfo{author}{Uriel \surnamestart Feige\surnameend},
  \bibinfo{author}{Abraham \surnamestart Flaxman\surnameend},
  \bibinfo{author}{Adam \surnamestart Kalai\surnameend}, \bibinfo{author}{Vahab
  \surnamestart Mirrokni\surnameend} \& \bibinfo{author}{Moshe \surnamestart
  Tennenholtz\surnameend} (\bibinfo{year}{2008}):
  \emph{\bibinfo{title}{Trust-based recommendation systems: an axiomatic
  approach}}.
\newblock In: {\sl \bibinfo{booktitle}{Proceedings of the 17th international
  conference on World Wide Web (WWW)}}, \bibinfo{address}{Beijing, China}, pp.
  \bibinfo{pages}{199--208}, \doi{10.1145/1367497.1367525}.

\bibitemdeclare{book}{Arr51}
\bibitem{Arr51}
\bibinfo{author}{Kenneth~Joseph \surnamestart Arrow\surnameend}
  (\bibinfo{year}{1951}): \emph{\bibinfo{title}{Social Choice and Individual
  Values}}.
\newblock \bibinfo{publisher}{Yale University Press}.

\bibitemdeclare{inproceedings}{AMT05}
\bibitem{AMT05}
\bibinfo{author}{Paolo \surnamestart Avesani\surnameend},
  \bibinfo{author}{Paolo \surnamestart Massa\surnameend} \&
  \bibinfo{author}{Roberto \surnamestart Tiella\surnameend}
  (\bibinfo{year}{2005}): \emph{\bibinfo{title}{A Trust-enhanced Recommender
  System Application: Moleskiing}}.
\newblock In: {\sl \bibinfo{booktitle}{Proceedings of the 20th ACM symposium on
  Applied computing (SAC)}}, \bibinfo{address}{Santa Fe, New Mexico}, pp.
  \bibinfo{pages}{1589--1593}, \doi{10.1145/1066677.1067036}.

\bibitemdeclare{inproceedings}{BL07}
\bibitem{BL07}
\bibinfo{author}{James \surnamestart Bennett\surnameend} \&
  \bibinfo{author}{Stan \surnamestart Lanning\surnameend}
  (\bibinfo{year}{2007}): \emph{\bibinfo{title}{The {N}etflix Prize}}.
\newblock In: {\sl \bibinfo{booktitle}{Proceedings of KDD Cup and Workshop}}.
\newblock
  \urlprefix\url{http://brettb.net/project/papers/2007%20The%20Netflix%20Prize.pdf}.

\bibitemdeclare{inproceedings}{BCTMT10}
\bibitem{BCTMT10}
\bibinfo{author}{Christian \surnamestart Borgs\surnameend},
  \bibinfo{author}{Jennifer \surnamestart Chayes\surnameend},
  \bibinfo{author}{Adam~Tauman \surnamestart Kalai\surnameend},
  \bibinfo{author}{Azarakhsh \surnamestart Malekian\surnameend} \&
  \bibinfo{author}{Moshe \surnamestart Tennenholtz\surnameend}
  (\bibinfo{year}{2010}): \emph{\bibinfo{title}{A novel approach to propagating
  distrust}}.
\newblock In: {\sl \bibinfo{booktitle}{Proceedings of the 6th Workshop on
  Internet and Network Economics (WINE)}}, \bibinfo{address}{Palo Alto,
  California}, pp. \bibinfo{pages}{87--105}, \doi{10.1007/978-3-642-17572-5_8}.

\bibitemdeclare{inproceedings}{GSOG09}
\bibitem{GSOG09}
\bibinfo{author}{Inma \surnamestart Garcia\surnameend}, \bibinfo{author}{Laura
  \surnamestart Sebastia\surnameend}, \bibinfo{author}{Eva \surnamestart
  Onaindia\surnameend} \& \bibinfo{author}{Cesar \surnamestart
  Guzman\surnameend} (\bibinfo{year}{2009}): \emph{\bibinfo{title}{A group
  recommender system for tourist activities}}.
\newblock In: {\sl \bibinfo{booktitle}{Proceedings of the 10th International
  Conference on E-Commerce and Web Technologies (EC-Web)}},
  \bibinfo{address}{Linz, Austria}, pp. \bibinfo{pages}{26--37},
  \doi{10.1007/978-3-642-03964-5_4}.

\bibitemdeclare{inproceedings}{GXLBHMS10}
\bibitem{GXLBHMS10}
\bibinfo{author}{Mike \surnamestart Gartrell\surnameend},
  \bibinfo{author}{Xinyu \surnamestart Xing\surnameend}, \bibinfo{author}{Qin
  \surnamestart Lv\surnameend}, \bibinfo{author}{Aaron \surnamestart
  Beach\surnameend}, \bibinfo{author}{Richard \surnamestart Han\surnameend},
  \bibinfo{author}{Shivakant \surnamestart Mishra\surnameend} \&
  \bibinfo{author}{Karim \surnamestart Seada\surnameend}
  (\bibinfo{year}{2010}): \emph{\bibinfo{title}{Enhancing group recommendation
  by incorporating social relationship interactions}}.
\newblock In: {\sl \bibinfo{booktitle}{Proceedings of the 16th ACM
  international conference on Supporting group work (GROUP)}}, pp.
  \bibinfo{pages}{97--106}, \doi{10.1145/1880071.1880087}.

\bibitemdeclare{inproceedings}{GLRW13}
\bibitem{GLRW13}
\bibinfo{author}{Jagadeesh \surnamestart Gorla\surnameend},
  \bibinfo{author}{Neal \surnamestart Lathia\surnameend},
  \bibinfo{author}{Stephen \surnamestart Robertson\surnameend} \&
  \bibinfo{author}{Jun \surnamestart Wang\surnameend} (\bibinfo{year}{2013}):
  \emph{\bibinfo{title}{Probabilistic group recommendation via information
  matching}}.
\newblock In: {\sl \bibinfo{booktitle}{Proceedings of the 22nd international
  conference on World Wide Web (WWW)}}, \bibinfo{address}{Seoul, Korea}, pp.
  \bibinfo{pages}{495--504}, \doi{10.1145/2488388.2488432}.

\bibitemdeclare{inproceedings}{GS09}
\bibitem{GS09}
\bibinfo{author}{Anil \surnamestart G\"{u}rsel\surnameend} \&
  \bibinfo{author}{Sandip \surnamestart Sen\surnameend} (\bibinfo{year}{2009}):
  \emph{\bibinfo{title}{Producing Timely Recommendations From Social Networks
  Through Targeted Search}}.
\newblock In: {\sl \bibinfo{booktitle}{Proceedings of The 8th International
  Conference on Autonomous Agents and Multiagent Systems (AAMAS)}},
  \bibinfo{volume}{2}, \bibinfo{address}{Budapest, Hungary}, pp.
  \bibinfo{pages}{805--812}, \doi{10.1145/1558109.1558123}.

\bibitemdeclare{inproceedings}{HS10}
\bibitem{HS10}
\bibinfo{author}{Chung \surnamestart wei Hang\surnameend} \&
  \bibinfo{author}{Munindar~P. \surnamestart Singh\surnameend}
  (\bibinfo{year}{2010}): \emph{\bibinfo{title}{Trust-Based Recommendation
  Based on Graph Similarity}}.
\newblock In: {\sl \bibinfo{booktitle}{Proceedings of the 13th International
  Workshop on Trust in Agent Societies}}, \bibinfo{address}{Toronto, Canada},
  pp. \bibinfo{pages}{71--81}.

\bibitemdeclare{article}{KKOR10}
\bibitem{KKOR10}
\bibinfo{author}{Jae~Kyeong \surnamestart Kim\surnameend},
  \bibinfo{author}{Hyea~Kyeong \surnamestart Kim\surnameend},
  \bibinfo{author}{Hee~Young \surnamestart Oh\surnameend} \&
  \bibinfo{author}{Young~U. \surnamestart Ryu\surnameend}
  (\bibinfo{year}{2010}): \emph{\bibinfo{title}{A group recommendation system
  for online communities}}.
\newblock {\sl \bibinfo{journal}{International Journal of Information
  Management}} \bibinfo{volume}{30}(\bibinfo{number}{3}), pp.
  \bibinfo{pages}{212--219}, \doi{10.1016/j.ijinfomgt.2009.09.006}.

\bibitemdeclare{inproceedings}{OCKR01}
\bibitem{OCKR01}
\bibinfo{author}{Mark \surnamestart O'Connor\surnameend}, \bibinfo{author}{Dan
  \surnamestart Cosley\surnameend}, \bibinfo{author}{Joseph~A. \surnamestart
  Konstan\surnameend} \& \bibinfo{author}{John \surnamestart Riedl\surnameend}
  (\bibinfo{year}{2001}): \emph{\bibinfo{title}{PolyLens: a recommender system
  for groups of users}}.
\newblock In: {\sl \bibinfo{booktitle}{Proceedings of the 7th conference on
  European Conference on Computer Supported Cooperative Work (ECSCW)}},
  \bibinfo{address}{Bonn, Germany}, pp. \bibinfo{pages}{199--218},
  \doi{10.1007/0-306-48019-0_11}.

\bibitemdeclare{inproceedings}{PMLR04}
\bibitem{PMLR04}
\bibinfo{author}{Jordi \surnamestart Palau\surnameend}, \bibinfo{author}{Miquel
  \surnamestart Montaner\surnameend}, \bibinfo{author}{Beatriz \surnamestart
  L{\'o}pez\surnameend} \& \bibinfo{author}{Josep~Llu{\'\i}s \surnamestart
  de~la Rosa\surnameend} (\bibinfo{year}{2004}):
  \emph{\bibinfo{title}{Collaboration Analysis in Recommender Systems Using
  Social Networks}}.
\newblock In: {\sl \bibinfo{booktitle}{Proceedings of 8th Cooperative
  Information Agents International Workshop (CIA)}}, {\sl
  \bibinfo{series}{Lecture Notes in Computer Science}} \bibinfo{volume}{3191},
  \bibinfo{address}{Erfurt, Germany}, pp. \bibinfo{pages}{137--151},
  \doi{10.1007/978-3-540-30104-2_11}.

\bibitemdeclare{inproceedings}{PHG00}
\bibitem{PHG00}
\bibinfo{author}{David~M. \surnamestart Pennock\surnameend},
  \bibinfo{author}{Eric \surnamestart Horvitz\surnameend} \&
  \bibinfo{author}{C.~Lee \surnamestart Giles\surnameend}
  (\bibinfo{year}{2000}): \emph{\bibinfo{title}{Social Choice Theory and
  Recommender Systems: Analysis of the Axiomatic Foundations of Collaborative
  Filtering}}.
\newblock In: {\sl \bibinfo{booktitle}{Proceedings of the 17th National
  Conference on Artificial Intelligence (AAAI)}}, \bibinfo{address}{Austin,
  Texas}, pp. \bibinfo{pages}{729--734}.

\bibitemdeclare{article}{Plo76}
\bibitem{Plo76}
\bibinfo{author}{Charles~R. \surnamestart Plott\surnameend}
  (\bibinfo{year}{1976}): \emph{\bibinfo{title}{Axiomatic Social Choice Theory:
  An Overview and Interpretation}}.
\newblock {\sl \bibinfo{journal}{American Journal of Political Science}}
  \bibinfo{volume}{20}(\bibinfo{number}{3}), pp. \bibinfo{pages}{511--596},
  \doi{10.2307/2110686}.

\bibitemdeclare{article}{RV97}
\bibitem{RV97}
\bibinfo{author}{Paul \surnamestart Resnick\surnameend} \&
  \bibinfo{author}{Hal~R. \surnamestart Varian\surnameend}
  (\bibinfo{year}{1997}): \emph{\bibinfo{title}{Recommender systems}}.
\newblock {\sl \bibinfo{journal}{Communications of the ACM}}
  \bibinfo{volume}{40}(\bibinfo{number}{3}), pp. \bibinfo{pages}{56--58},
  \doi{10.1145/245108.245121}.

\bibitemdeclare{incollection}{RZ02}
\bibitem{RZ02}
\bibinfo{author}{Paul \surnamestart Resnick\surnameend} \&
  \bibinfo{author}{Richard \surnamestart Zeckhauser\surnameend}
  (\bibinfo{year}{2002}): \emph{\bibinfo{title}{Trust Among Strangers in
  Internet Transactions: Empirical Analysis of e{B}ay's Reputation System}}.
\newblock In \bibinfo{editor}{Michael~R. \surnamestart Baye\surnameend},
  editor: {\sl \bibinfo{booktitle}{The Economics of the Internet and E-commerce
  (Advances in Applied Microeconomics)}}, \bibinfo{volume}{11},
  \bibinfo{publisher}{Emerald Group Publishing Limited}, pp.
  \bibinfo{pages}{127--157}, \doi{10.1016/S0278-0984(02)11030-3}.

\bibitemdeclare{inproceedings}{RT09}
\bibitem{RT09}
\bibinfo{author}{Ola \surnamestart Rozenfeld\surnameend} \&
  \bibinfo{author}{Moshe \surnamestart Tennenholtz\surnameend}
  (\bibinfo{year}{2009}): \emph{\bibinfo{title}{Consistent Continuous
  Trust-Based Recommendation Systems}}.
\newblock In: {\sl \bibinfo{booktitle}{Proceedings of the 5th Workshop on
  Internet and Network Economics (WINE)}}, \bibinfo{address}{Rome, Italy}, pp.
  \bibinfo{pages}{113--124}, \doi{10.1007/978-3-642-10841-9_12}.

\bibitemdeclare{inproceedings}{SKR99}
\bibitem{SKR99}
\bibinfo{author}{J.~Ben \surnamestart Schafer\surnameend},
  \bibinfo{author}{Joseph \surnamestart Konstan\surnameend} \&
  \bibinfo{author}{John \surnamestart Riedi\surnameend} (\bibinfo{year}{1999}):
  \emph{\bibinfo{title}{Recommender Systems in E-Commerce}}.
\newblock In: {\sl \bibinfo{booktitle}{Proceedings of the 1st ACM conference on
  Electronic Commerce (EC)}}, \bibinfo{address}{Denver, Colorado}, pp.
  \bibinfo{pages}{158--166}, \doi{10.1145/336992.337035}.

\bibitemdeclare{inproceedings}{SYMM11}
\bibitem{SYMM11}
\bibinfo{author}{Shunichi \surnamestart Seko\surnameend},
  \bibinfo{author}{Takashi \surnamestart Yagi\surnameend},
  \bibinfo{author}{Manabu \surnamestart Motegi\surnameend} \&
  \bibinfo{author}{Shinyo \surnamestart Muto\surnameend}
  (\bibinfo{year}{2011}): \emph{\bibinfo{title}{Group recommendation using
  feature space representing behavioral tendency and power balance among
  members}}.
\newblock In: {\sl \bibinfo{booktitle}{Proceedings of the 5th ACM conference on
  Recommender systems (RecSys)}}, \bibinfo{address}{Chicago, Illinois}, pp.
  \bibinfo{pages}{101--108}, \doi{10.1145/2043932.2043953}.

\bibitemdeclare{inproceedings}{Mos04}
\bibitem{Mos04}
\bibinfo{author}{Moshe \surnamestart Tennenholtz\surnameend}
  (\bibinfo{year}{2004}): \emph{\bibinfo{title}{Reputation systems: an
  axiomatic approach}}.
\newblock In: {\sl \bibinfo{booktitle}{Proceedings of the 20th conference on
  Uncertainty in Artificial Intelligence (UAI)}}, \bibinfo{address}{Banff,
  Canada}, pp. \bibinfo{pages}{544--551}.

\bibitemdeclare{article}{WBS08}
\bibitem{WBS08}
\bibinfo{author}{Frank~Edward \surnamestart Walter\surnameend},
  \bibinfo{author}{Stefano \surnamestart Battiston\surnameend} \&
  \bibinfo{author}{Frank \surnamestart Schweitzer\surnameend}
  (\bibinfo{year}{2008}): \emph{\bibinfo{title}{A model of a trust-based
  recommendation system on a social network}}.
\newblock {\sl \bibinfo{journal}{Journal of Autonomous Agents and Multi-Agent
  Systems (JAAMAS)}} \bibinfo{volume}{16}(\bibinfo{number}{1}), pp.
  \bibinfo{pages}{57--74}, \doi{10.1007/s10458-007-9021-x}.

\bibitemdeclare{inproceedings}{YCL14}
\bibitem{YCL14}
\bibinfo{author}{Quan \surnamestart Yuan\surnameend}, \bibinfo{author}{Gao
  \surnamestart Cong\surnameend} \& \bibinfo{author}{Chin-Yew \surnamestart
  Lin\surnameend} (\bibinfo{year}{2014}): \emph{\bibinfo{title}{COM: a
  generative model for group recommendation}}.
\newblock In: {\sl \bibinfo{booktitle}{Proceedings of the 20th ACM SIGKDD
  international conference on Knowledge discovery and data mining (KDD)}},
  \bibinfo{address}{New York City, New York}, pp. \bibinfo{pages}{163--172},
  \doi{10.1145/2623330.2623616}.

\bibitemdeclare{techreport}{ZGMZG15}
\bibitem{ZGMZG15}
\bibinfo{author}{Cheng \surnamestart Zhang\surnameend}, \bibinfo{author}{Mike
  \surnamestart Gartrell\surnameend}, \bibinfo{author}{Thomas~P. \surnamestart
  Minka\surnameend}, \bibinfo{author}{Yordan \surnamestart Zaykov\surnameend}
  \& \bibinfo{author}{John \surnamestart Guiver\surnameend}
  (\bibinfo{year}{2015}): \emph{\bibinfo{title}{GroupBox: A generative model
  for group recommendation}}.
\newblock \bibinfo{type}{Technical Report} \bibinfo{number}{MSR-TR--2015-61},
  \bibinfo{institution}{Microsoft Research}.

\end{thebibliography}
%
%

\end{document}